\documentclass[a4paper,10pt]{article}
\usepackage[utf8]{inputenc}
\usepackage[title]{appendix}
\usepackage{enumerate}
\usepackage{amssymb, amsmath, amsthm}
\usepackage{graphicx}
\usepackage{listings}
\usepackage{hyperref}
\usepackage{xcolor}
\makeatletter
\renewcommand*{\eqref}[1]{%
  \hyperref[{#1}]{\textup{\tagform@{\ref*{#1}}}}%
}
\makeatother
\usepackage{mathtools}

\AtBeginDocument{\renewcommand{\d}{\textrm{d}}}
\newcommand{\cc}{^*}
\newcommand{\abs}[1]{\left| #1 \right|}
\newcommand{\R}{\mathbb{R}} 
\newcommand{\N}{\mathbb{N}} 
\newcommand{\lt}{\left}
\newcommand{\rt}{\right}
\newcommand{\eg}{\textit{e.g.}}

\newenvironment{definition}[1][Definition]{\begin{trivlist}
\item[\hskip \labelsep {\bfseries #1}]}{\end{trivlist}}

\newenvironment{example}[1][Example]{\begin{trivlist}
\item[\hskip \labelsep {\bfseries #1}]}{\end{trivlist}}

\mathtoolsset{showonlyrefs=true}

\title{On the nonintegrability of equations for long- and short-wave interactions}
\author{Bernard Deconinck and Jeremy Upsal\\
  \\
Department of Applied Mathematics,\\
University of Washington,\\
Seattle, WA 98195, USA}

\begin{document}

\maketitle

\begin{abstract}
We examine the integrability of two models used for the interaction of long and
  short waves in dispersive media. One is more classical but arguably cannot be
  derived from the underlying water wave equations, while the other one was
  recently derived. We use the method of Zakharov and Schulman to attempt to
  construct conserved quantities for these systems at different orders in the
  magnitude of the solutions. The coupled KdV-NLS model is shown to be
  nonintegrable, due to the presence of fourth-order resonances. A coupled real
  KdV - complex KdV system is shown to suffer the same fate, except for three
  special choices of the coefficients, where higher-order calculations or a
  different approach are necessary to conclude integrability or the absence
  thereof. 
\end{abstract}

\section{Introduction}
Systems that couple long and short waves have generated significant interest
recently (\eg ~\cite{MR3129020, MR2240820, MR1681850, MR2309177, MR2674102}).
Much attention in this area has been devoted to the following system, known as
the cubic nonlinear Schr\"{o}dinger-Korteweg-deVries (NLS-KdV) system:
\begin{align}
\begin{split}
 i u_t + u_{xx} + \alpha \abs{u}^2 u &= -\beta uv,\\
 v_t +\gamma v v_x + v_{xxx} &= - \beta(\abs{u}^2)_x,
\end{split}\label{eqn:NLS-KdV}
\end{align}
where $\alpha,\beta$ and $\gamma$ are real constants, $x \in \R$, $v$ is
a real-valued function, and $u$ is a complex-valued function. Recently, it was
shown that \eqref{eqn:NLS-KdV} cannot be consistently derived starting from the
underlying water wave equations \cite{deconinckSegal-LongShort}. The following
coupled KdV-CKdV (Complex KdV) model was suggested as an alternative with a
consistent derivation:
\begin{align}
  \begin{split} u_t + 2\beta u_x + \alpha u_{xxx} &= -
    2\beta(u v)_x ,\\ v_t + \beta v_x + \beta v v_x + \gamma v_{xxx} &= -
    \beta(|u|^2)_x .
  \end{split} \label{eqn:consistent}
\end{align}
As above, $v~(u)$ is a real- (complex-) valued function and $\alpha,\beta$ and
$\gamma$ are real constants. The coefficients in the system above occur as they
are by using the scaling symmetries of the system to minimize the number of
free parameters \cite{deconinckSegal-LongShort}. We examine whether or not the
two systems \eqref{eqn:NLS-KdV} and \eqref{eqn:consistent} are integrable in a
sense detailed below.

A method for showing the nonintegrability of a system developed by Zakharov and
Schulman \cite{zakharovSchulmanImportantDevelopmentsInScattering,
zakharovSchulmanWhatIsIntegrability} distinguishes between \textit{completely
integrable systems} and \textit{solvable systems}. Completely integrable
systems are those for which we can find action-angle variables and solvable
equations are those which can be solved by the inverse scattering transform
(IST) \cite{ablowitz_segur-IST}. Since integrability is a feature of the
equations and not of a particular solution, we may always assume that we are
working in a neighborhood of a solution with a nondegenerate linearization. 

The test for complete integrability has the following steps:
\begin{enumerate}
  \item Any completely integrable Hamiltonian system may be written locally in
      action-angle variables.
  \item A system in action-angle variables is equivalent to a collection of
    uncoupled harmonic oscillators, so its Hamiltonian is quadratic.
  \item Near-identity normal-form transformations \cite{Wiggins} can be used to
    reduce any Hamiltonian to quadratic as long as there are no obstructions
    from resonances.
  \item Any obstruction in the above steps due to resonances implies the system
      is not completely integrable.
\end{enumerate}
The normal-form transformation that removes $n$-th order terms from the
Hamiltonian gives rise to a resonance manifold which describes the process of
scattering $p$ waves ($p \in \N$) into $n-p$ waves. For example, if a system
admits two dispersion laws $\omega^{(1)}$ and $\omega^{(2)}$, an $n$-th order
resonance manifold is defined by
\begin{align}
  M = \lt\{(k_1,\ldots,k_n)\in \R^n \bigg| \sum_{j=1}^n \sigma_j k_j = 0 \text{ and }
  \sum_{j=1}^n \sigma_j\omega^{(\ell)}(k_j) = 0\rt\}, 
\label{defn:resonanceManifold}
\end{align}
with any combination of $\sigma_j \in \{-1,1\}$ and $\ell\in\{1,2\}$.
Associated with each resonance manifold is an interaction coefficient function
which describes the amplitude of the scattering process. If the coefficient
function vanishes on the resonance manifold then the singularity of the normal
form transformation is removable and the transformation is valid. If the
coefficient function does not vanish on the resonance manifold, complete
integrability is not possible but solvability may be.

\vspace{0.1in}

The test for solvability has the following steps:
\begin{enumerate}
  \item Every system solvable by the IST has an infinite hierarchy of equations
      solvable by the IST. The members of the hierarchy share conserved
      quantities.
  \item By assumption, any equation solvable by the IST is linearizable with
      nondegenerate linearization, so each member of the hierarchy has
      quadratic terms in the Hamiltonian, at least in the small amplitude
      limit.
  \item Every member of the hierarchy has a linearly independent Hamiltonian, so
    the original system has infinitely many conserved quantities with linearly
    independent quadratic terms (see \eg \cite{magri1978Hamiltonian}).
  \item If there exist only finitely many conserved quantities with quadratic
    terms for our PDE, it is not solvable by the IST.
\end{enumerate}
The method of Zakharov and Schulman begins by removing all higher-order
nonresonant terms as above. Next an ansatz is made about the existence of an
additional conserved quantity in a power series in terms of unknown amplitudes.
Upon enforcing that the quantity is independent of $t$, resonance manifolds
appear as above. However, in this case, the resonance manifold coefficient
function is multiplied by another quantity:
\begin{align}
  \sum_{j=1}^n \sigma_j \Phi^{(\ell)}(k_j), 
  \label{eqn:degeneracyCondition}
\end{align}
where $\sigma_j$ and $\ell$ are the same as in \eqref{defn:resonanceManifold}
and $\Phi^{(\ell)}$ are the unknown quadratic amplitudes in the power series.
If functions $\Phi^{(\ell)}$, linearly independent from the two relations
defining the resonance manifold, can be found such that
\eqref{eqn:degeneracyCondition} equals zero, then the manifold is called
degenerate
\cite{zakharov_schulman-degenerativeDispersionLaws}. If any of the $n$-th order
resonance manifolds are nondegenerate and have nonzero coefficient function,
the constructed quantity is not conserved. The fact that another conserved
quantity with linearly independent quadratic terms cannot be constructed
implies that the system must not be solvable by the IST.

Determining whether or not a resonance manifold is degenerate poses challenges.
We use the theory of web geometry \cite{balk} to check degeneracy as described
in Appendix \ref{sec:webGeometry}. In Sections \ref{sec:NLS-KdV} and
\ref{sec:consistent} we examine the integrability of \eqref{eqn:NLS-KdV} and
\eqref{eqn:consistent}.

\section{Coupled NLS \& KdV Model \label{sec:NLS-KdV}}
The Hamiltonian for \eqref{eqn:NLS-KdV} on the whole line is
\begin{align}
 H = \int \lt(|u_x|^2 + \frac12 v_x^2 - \frac{\alpha}{2}|u|^4 - \frac{\gamma}{6}
 v^3 - \beta|u|^2 v\rt)\d x\label{ham:NLS-KdV},
\end{align}
for the variables $z = (u, i u\cc, v)$ with non-canonical Poisson structure
\[
 J = \begin{pmatrix}
      0 & 1 & 0\\
      -1 & 0 & 0\\
      0 & 0 & \partial_x
     \end{pmatrix},
\]
so that \eqref{eqn:NLS-KdV} is equivalent to $z_t = J \delta H/\delta z$, where
$\delta/\delta z$ denotes the variational gradient with respect to the
components of $z$ \cite{Fordy_Antonowicz-HamStruct}. Here and throughout
integrals without bounds are to be interpreted as whole line integrals. This
system admits two types of waves with dispersion relations $\omega_k = k^2$ and
$\Omega_k = -k^3$. Here and throughout, $k$ subscripts are indices, not partial
derivatives. We introduce the Fourier transform,
\begin{align}
  u(x) = \frac1{\sqrt{2\pi}} \int u(k)e^{ikx}\d k = \frac{1}{\sqrt{2\pi}} \int
  u_k e^{ikx} \d k. \label{FourierTransform} 
\end{align}
Applying the Fourier transform to $u$ and $v$ results in a Hamiltonian
system for $(u_k, v_k)$ with Hamiltonian
\begin{align}
  \begin{split}
 H(u_k,v_k) &= \int k^2 u_k u_k\cc \d k + \int_0^\infty k^2v_k v_k\cc \d k -
 \frac{\beta}{\sqrt{2\pi}} \int u_1\cc v_2 u_3 \delta_{1-2-3}\d_{123} \\
 &\qquad -\frac{\gamma}{6\sqrt{2\pi}}\int v_1 v_2 v_3 \delta_{123}\d_{123} 
 - \frac{\alpha}{2(2\pi)}\int u_1 u_2 u_3\cc u_4\cc
 \delta_{12-3-4}\d_{1234},
 \end{split}\label{NLS-KdVHamiltonian}
\end{align}
where we use the notation $u_j = u_{k_j}, \d_{123} = \d k_1 \d
k_2 \d k_3$, $u_k\cc$ denotes the complex conjugate of $u_k$, and $\delta_{12-3}
= \delta(k_1 + k_2 - k_3)$ where $\delta(\cdot)$ is the Dirac-delta function.
The integral with quadratic integrand in $v_k$ found in
\eqref{NLS-KdVHamiltonian} is reduced to an integral on the half-line using the
fact that $v_k\cc = v_{-k}$ since $v(x)$ is real. In Fourier variables, the
dynamics are 
\begin{align}
 i \dot u_k = \frac{\delta H}{\delta u_k\cc}, \qquad \dot v_k = ik\frac{\delta
 H}{\delta v_k\cc}.
\end{align}
We introduce $a_k$ by
\begin{align}
 v_k = |k|^{1/2}(a_k \theta_{-k} + a_{-k}\cc \theta_{k}),
\end{align}
where 
\begin{align}
  \theta_k = \theta(k) = \begin{cases}
    0,& k < 0, \\
                        1, & k\geq 0 ,
                      \end{cases}
\end{align}
is the Heaviside-function. The dynamics are 
\begin{align}
 i \dot u_k = \frac{\delta H}{\delta u_k\cc}, \qquad i \dot a_k = \frac{\delta
 H}{\delta a_k\cc},\label{dynamics:NLS-KdV}
\end{align}
with
\begin{align}
\begin{split}
 H(u_k, a_k) &= H_2(u_k,a_k) + H_3(u_k,a_k) + H_4(u_k,a_k),\\ 
 H_2(u_k,a_k) &= \int \omega_k u_k u_k\cc \d k + \int_{-\infty}^0
 \Omega_k a_k a_k\cc \d k,\\ 
 H_3 (u_k,a_k)&= \int U_{123}(a_1\cc a_2 a_3 + a_1 a_2\cc
 a_3\cc)\delta_{1-2-3}\d_{123} + \int V_{123} (u_1\cc a_2 u_3 + u_1 a_2\cc
 u_3\cc)\delta_{1-2-3}\d_{123},\\
 H_4 (u_k,a_k)&= \int W_{1234} u_1 u_2 u_3\cc
 u_4\cc\delta_{12-3-4}\d_{1234},\\ 
 U_{123} &= -\frac{\gamma}{2\sqrt{2\pi}}\abs{k_1
 k_2k_3}^{1/2}\theta_{-1}\theta_{-2}\theta_{-3},
 \quad V_{123} = -\frac{\beta}{\sqrt{2 \pi}} \abs{k_2}^{1/2} \theta_{-2}, \quad
 W_{1234} = -\frac{\alpha}{2(2\pi)}.
\end{split}\label{NLS-KdV:Hamiltonian}
\end{align}
Up to third-order, this system is identical to that used to study the
integrability of Langmuir Waves \cite{benilov1983integrability} with the $k$
there replaced by $-k$ here. The canonical near-identity transformation,
\begin{align}
\begin{split}
 a_k &= \tilde a_k + \int\lt(U_{012}^{(1)} \tilde a_1 \tilde a_2 -2
 U_{120}^{(1)} \tilde a_1 \tilde a_2\cc - U_{102}^{(2)} \tilde u_1 \tilde
 u_2\cc\rt)\d_{12},\\ 
 u_k &= \tilde u_k + \int \lt(U_{012}^{(2)} \tilde a_1 \tilde
 u_2 - U^{(2)}_{210} \tilde a_1 \cc \tilde u_2\rt)\d_{12},\\ 
 U_{\ell m n}^{(1)} &= -\frac{U_{\ell m n}}{\Omega_\ell - \Omega_m -
 \Omega_n}\delta_{\ell - m - n},
 \qquad U_{\ell m n}^{(2)} = - \frac{V_{\ell m n}}{\omega_\ell - \Omega_m -
 \omega_n}\delta_{\ell-m-n},
\end{split}
\label{NLS-KdV:ThirdOrderRemoval}
\end{align}
removes third-order terms from the Hamiltonian so that
$H(\tilde u_k, \tilde a_k) = H_2(\tilde u_k, \tilde a_k) + H_4(\tilde u_k,
\tilde a_k) + \tilde H_4(\tilde u_k, \tilde a_k)$ where
$H_2$ is unchanged from \eqref{NLS-KdV:Hamiltonian} and $\tilde H_4$ are the
quartic terms which arise from $H_3$ under \eqref{NLS-KdV:ThirdOrderRemoval}.

The transformation \eqref{NLS-KdV:ThirdOrderRemoval} gives rise to two resonance
manifolds,
\begin{align}
  M_1 = \lt\{ (k_1,k_2,k_3)\in \R^3 : k_1 - k_2 - k_3 = 0 \text{ and } \Omega(k_1) -
  \Omega(k_2) -\Omega(k_3) = 0 \rt\}, \\
  M_2 = \lt\{(k_1,k_2,k_3) \in \R^3 :
    k_1 - k_2 - k_3 = 0 \text{ and } \omega(k_1) - \Omega(k_2) - \omega(k_3) = 0
  \rt\}.
\end{align}
The coefficient function $U_{123} = 0$ on $M_1$ and $V_{123} = 0$ on $M_2$ so the
singularities in $U_{123}^{(1)}$ and $U_{123}^{(2)}$ are removable. Next we
seek to remove the fourth-order terms from the Hamiltonian using a
near-identity transformation. One resonance manifold appearing in such a
transformation is defined by
\begin{align}
  M_3 = \lt\{(k_1,k_2,k_3,k_4)\in \R^4 : k_1 + k_2 - k_3 - k_4 = 0 \text{ and } \omega(k_1) + \Omega(k_2) -
  \omega(k_3) - \Omega(k_4) = 0\rt\},\label{NLS-KdV:fourWaveProcess}
\end{align}
corresponding to the process of converting two waves $k_1$ and $k_2$ with
frequency $\omega(k_1)$ and $\Omega(k_2)$ respectively to two with frequency
$\omega(k_3)$ and $\Omega(k_4)$. The coefficient function of this process is
found by collecting the fourth-order terms which multiply the quantity
$\delta_{12-3-4}/(\omega_1+\Omega_2-\omega_3-\Omega_4)$:
\begin{align*}
 T_{k_1, k_2,k_3,k_4} &= T^{(1)}_{k_1, k_2, k_3, k_4} +
 T^{(2)}_{k_1,k_2,k_3,k_4}, \\
 T^{(1)}_{k_1,k_2,k_3,k_4} &= 2 \lt( \frac{V_{k_3 +
 k_4,k_4,k_3}V_{k_1+k_2,k_2,k_1}}{\omega_{k_1} + \Omega_{k_2} -
 \omega_{k_1+k_2}} + \frac{V_{k_1,k_4, k_1-k_4}
 V_{k_3,k_2,k_3-k_2}}{\omega_{k_1} - \Omega_{k_4} - \omega_{k_1-k_4}}\rt)\\
 &\qquad + 4 \lt( \frac{ V_{k_1,k_1-k_3,k_3}U_{k_4,k_2,k_4-k_2}}{\Omega_{k_4} -
 \Omega_{k_2} - \Omega_{k_4 - k_2}}  + \frac{ V_{k_3,
 k_3-k_1,k_1}U_{k_4,k_2,k_2-k_4}}{\Omega_{k_2}-\Omega_{k_4} -
 \Omega_{k_2-k_4}}\rt),\\
 T^{(2)}_{k_1,k_2,k_3,k_4} &=  \frac{\omega_{k_1+k_2}V_{k_4 +k_3, k_4, k_3}
 V_{k_1+k_2, k_2, k_1}}{(\omega_{k_4 + k_3}-\Omega_{k_4} -
 \omega_{k_3})(\omega_{k_2 + k_1} - \Omega_2 - \omega_1)} + \frac{\omega_{k_3 -
 k_2} V_{k_3, k_2, k_3-k_2} V_{k_1,k_4,k_1-k_4}}{(\omega_{k_3} - \Omega_{k_2} -
 \omega_{k_3-k_2})(\omega_{k_1} - \Omega_{k_4} - \omega_{k_1-k_4})}\\
 &\quad + 2\frac{\Omega_{k_4-k_2} U_{k_4,k_2,k_4-k_2}
 V_{k_1,k_1-k_3,k_3}}{(\Omega_{k_4} - \Omega_{k_2} -
 \Omega_{k_4-k_2})(\omega_{k_1} - \Omega_{k_1-k_3} - \omega_{k_3})} +
 2\frac{\Omega_{k_2-k_4}U_{k_2,k_4,k_2-k_4} V_{k_3,k_3-k_1,k_1}}{(\Omega_{k_2} -
 \Omega_{k_4} - \Omega_{k_2-k_4})(\omega_3 - \Omega_{k_3 - k_1} -
 \omega_{k_1})}. 
\end{align*}
The quantity $T_{k_1,k_2,k_3,k_4}^{(1)}$ is (10) in \cite{benilov1983integrability}
but $T_{k_1,k_2,k_3,k_4}^{(2)}$ is mistakenly omitted from the full expression
for $T_{k_1,k_2,k_3,k_4}$ \cite{benilovPersonal}. The interaction coefficient
$T_{k_1,k_2,k_3,k_4}^{(2)}$ originates from the product of the two quadratic
terms of the transformation (\ref{NLS-KdV:ThirdOrderRemoval}) when applied to
the quadratic part of the Hamiltonian, $H_2$. The result of
\cite{benilov1983integrability} remains unchanged, $T_{k_1,k_2,k_3,k_4}$ is not
identically zero on $M_3$.

The scattering process defined by the resonance manifold $M_3$ is proven to be
nondegenerate in \cite{benilov1983integrability} using elementary methods. Here
we use web geometry (Appendix \ref{sec:webGeometry}) to show nondegeneracy since
this technique generalizes in a much more straightforward manner and will be
used later for studying \eqref{eqn:consistent}. We define a family of foliations
of $M_3$ by
\begin{align}
  k_j = \text{constant}, \quad j = 1,2,3,4,\label{foliations}
\end{align}
which is a 4-web of $M_3$. The process \eqref{NLS-KdV:fourWaveProcess} does not
correspond to billiard scattering since $M_3$ can be parameterized by
\begin{align}
  k_1 &= \frac12\lt(-k_2^2 - k_2k_4 - k_4^2+ k_4 - k_2\rt), \quad k_3 =
  \frac12\lt(-k_2^2 - k_2k_4 - k_4^2 + k_2 - k_4\rt).
  \label{NLS-KdVParameterization}
\end{align}
We use Mathematica to calculate the invariants introduced in
\cite{akivis2005linearizability} to show that this 4-web is linearizable only if
$\beta = 0$. However $\beta=0$ corresponds to an uncoupled system of KdV and NLS
equations and is known to be integrable. Since the web is not linearizable it
must have rank 2, hence this process is not degenerate.

Since there exists a fourth-order resonance manifold with nonzero interaction
coefficient, fourth-order terms cannot be removed from the Hamiltonian, thus
the system is not completely integrable. Since this resonance manifold is also
nondegenerate, a new conserved quantity cannot be constructed with linearly
independent quadratic part, so the system must not be solvable by the IST.
Equation \eqref{eqn:NLS-KdV} is nonintegrable in either sense defined in
Section 1.

\section{Coupled KdV-CKdV Model \label{sec:consistent}} 
The Hamiltonian for \eqref{eqn:consistent} on the whole line is 
\begin{align}
 H &= \int \lt( \frac{\alpha }{2} |u_x|^2 + \frac{\gamma}{2}v_x^2 -
 \frac{\beta}{6}v^3 - \beta|u|^2 v - \beta|u|^2 - \frac \beta2 v^2\rt) \d x,
\end{align}
for the variables $(u, iu\cc, v)$. The dynamics are
\begin{align*}
 u_t &= 2\partial_x \frac{\delta H}{\delta u\cc}, \qquad v_t = \partial_x \frac{\delta H}{\delta v}. 
\end{align*}
Equation \eqref{eqn:consistent} admits two types of waves with frequencies
$\omega_k = 2\beta k - \alpha k^3$ and $\Omega_k = \beta k - \gamma k^3$.
Applying the Fourier transform \eqref{FourierTransform} to $u$ and $v$ results
in a Hamiltonian system for $(u_k,v_k)$ with Hamiltonian
\begin{align}
 H(u_k, v_k) &= \frac{1}{2} \int (\alpha k^2 - 2\beta) u_k u_k\cc \d k + \int_0^\infty
 (\gamma k^2 -\beta) v_k v_k\cc \d k - \frac{\beta}{6\sqrt{2\pi}} \int v_1 v_2 v_3
 \delta_{123}\d_{123} \\
 &\qquad - \frac{\beta}{\sqrt{2\pi}} \int u_1\cc v_2 u_3
 \delta_{1-2-3}\d_{123}, \\
\end{align}
and dynamics 
\begin{align}
 \dot u_k &= 2ik \frac{\delta H}{\delta u_k\cc},\qquad \dot v_k = ik \frac{\delta H}{\delta v_k\cc}.
\end{align}
Here, as in \eqref{NLS-KdVHamiltonian}, the integral with quadratic integrand
in $v_k$ is reduced to a half-line integral using the reality of $v(x)$. 

Introducing the variables $a_k$ and $b_k$ by 
\begin{align}
 v_k &= |k|^{1/2}(a_k \theta_{-k} + a_{-k}\cc \theta_k), \qquad u_k = |2k|^{1/2} b_{-k}\cc,
\end{align}
the dynamical equations are rewritten as
\begin{align}
 i \dot a_k &= \frac{\delta H}{\delta a_k\cc}, \qquad i \dot b_k = \frac{\delta
 H}{\delta
 b_k\cc}, \qquad -i \dot b_{-k} = \frac{\delta H}{\delta b_{-k}\cc},
 \label{consistentDynamicalEqns}
\end{align}
with
\begin{align}
  \begin{split}
  H(a_k,b_k, b_{-k}) &= H_2(a_k,b_k,b_{-k}) + H_3(a_k,b_k,b_{-k}),\\
  H_2(a_k,b_k,b_{-k}) &= \int_{-\infty}^0 \Omega_k a_k a_k\cc \d k + \int_{-\infty}^0 \omega_k b_k
 b_k\cc \d k + \int_{-\infty}^0 \omega_k b_{-k} b_{-k}\cc \d k,\\
 H_3(a_k,b_k,b_{-k}) &= \int U_{123}\lt(b_{-1}a_2 b_{-3}\cc \delta_{1-2-3} + b_1\cc a_2 b_{-3}
 \delta_{1-23} + b_1\cc a_2 b_3\delta_{1-2-3}\rt)\d_{123} + c.c.\\
 &\qquad + \int V_{123} a_1\cc a_2 a_3 \delta_{1-2-3}\d_{123} + c.c.,\\
 V_{123} &= -\frac {\beta}{2\sqrt{2\pi}} |k_1k_2k_3|^{1/2}
 \theta_{-1}\theta_{-2}\theta_{-3}, \qquad U_{123} = 4 V_{123},
 \end{split}\label{consistentHamiltonian}
\end{align}
where $c.c.$ is the complex conjugate of the preceding terms. The dynamical
equations \eqref{consistentDynamicalEqns} are to be interpreted for $k<0$ only.
Half line integrals are used so that our system is in normal Hamiltonian
variables with the quadratic terms of the Hamiltonian being multiplied by the
frequencies \cite{zakharov1974hamiltonianFormalism}.

This system is Hamiltonian with canonical variables $(ia_k, a_k\cc, i b_k,
b_k\cc, ib_{-k}\cc, b_{-k})$ and canonical Poisson structure
\[
 J = \begin{pmatrix} 
      J_1 & 0 & 0\\
      0 & J_1 & 0\\
      0 & 0 & J_1
     \end{pmatrix}, \qquad J_1 = \begin{pmatrix} 0 & 1\\ -1 & 0 \end{pmatrix}.
\]

The canonical near-identity transformation to the variables $(i \tilde a_k,
\tilde a_k\cc, i \tilde b_k, \tilde b_k\cc, i \tilde b_{-k}\cc, \tilde b_{-k})$
given by
\begin{align}
 a_k &= \tilde a_k + \int \lt(\int A_{012}^{(1)} \tilde a_1 \tilde a_2 -
 2A_{120}^{(1)} \tilde a_1 \tilde a_2\cc - B_{102}^{(2)} \tilde b_{-1}\cc\tilde
 b_{-2} - B_{102}^{(1)} \tilde b_1 \tilde b_{-2}\cc - B_{102}^{(2)}\tilde b_1
 \tilde b_2\cc\rt)\d_{12},\\
 b_k &= \tilde b_k + \int\lt( B_{012}^{(1)} \tilde a_1 \tilde b_{-2} +
 B_{012}^{(2)}\tilde a_1 \tilde b_2 - B_{120}^{(2)}\tilde b_1 \tilde
 a_2\cc\rt)\d_{12},\label{consistentThirdOrderTermRemoval}\\
 b_{-k} &= \tilde b_{-k} + \int\lt(-B_{120}^{(2)} \tilde b_{-1} \tilde a_2 +
 B_{012}^{(2)}\tilde a_1\cc \tilde b_{-2} + B_{120}^{(1)} \tilde b_1 \tilde
 a_2\cc\rt)\d_{12},\\
 A_{\ell m n}^{(1)} &= -\frac{V_{\ell m n}}{\Omega_\ell -\Omega_m - \Omega_n}
 \delta_{\ell-m-n},
 \quad 
 B_{\ell m n}^{(1)} = - \frac{U_{\ell m n}}{\omega_\ell - \Omega_m + \omega_n}
 \delta_{\ell-mn},\quad
 B_{\ell m n}^{(2)} = - \frac{U_{\ell m n}}{\omega_\ell - \Omega_m -
 \omega_n}\delta_{\ell-m-n},
\end{align}
removes third-order terms from the Hamiltonian. The transformation
\eqref{consistentThirdOrderTermRemoval} gives rise to three separate three-wave
resonance manifolds:
\begin{align}
  M_1 &= \lt\{ (k_1,k_2,k_3)\in\R^3 : k_1 - k_2 - k_3 = 0 \text{ and } \Omega(k_1) -
  \Omega(k_2) - \Omega(k_3) = 0 \rt\},\\
  M_2 &= \lt\{ (k_1,k_2,k_3)\in\R^3 : k_1 - k_2 + k_3 = 0 \text{ and } \omega(k_1) -
  \Omega(k_2) + \omega(k_3) = 0 \rt\},\\
  M_3 &= \lt\{ (k_1,k_2,k_3)\in\R^3 : k_1 - k_2 - k_3 = 0 \text{ and } \omega(k_1) -
  \Omega(k_2) - \omega(k_3) = 0 \rt\}.
\end{align}
Since the coefficient $U_{123}$ vanishes on $M_2$ and $M_3$, $B_{123}^{(1)}$ and
$B_{123}^{(2)}$ have removable singularities only. Further, the coefficient
$V_{123}=0$ on $M_1$ unless $\gamma = 0$. However, the process defining $M_1$
is degenerate since all three-wave interaction processes in one-dimension are
degenerate \cite{zakharovSchulmanImportantDevelopmentsInScattering}. Thus we
must try to remove fourth-order terms. 

One resonance manifold which appears when attempting to remove fourth-order
terms from $H$ is defined by
\begin{align}
  {\cal M}_1 = \{(k_1,k_2,k_3,k_4) : k_1 = k_2 + k_3 + k_4 \text{ and } \omega_1 =
\Omega_2 + \omega_3 + \Omega_4\}. 
\end{align}
This manifold splits into two components with local coordinates,
\begin{align}
\begin{split}
 k_1 &= \frac 12(k_2+k_4) \pm \frac{1}{2\sqrt{3\alpha}}\lt[4\beta - (k_2^2 +
 k_4^2)(\alpha - 4\gamma) - 2k_2 k_4(\alpha + 2\gamma)\rt]^{1/2},\\
 k_3 &= -\frac 12(k_2+k_4) \pm \frac{1}{2\sqrt{3\alpha}}\lt[4\beta - (k_2^2 +
 k_4^2)(\alpha - 4\gamma) - 2k_2 k_4(\alpha + 2\gamma)\rt]^{1/2},
\end{split}
\label{consistentManifoldParameterization}
\end{align}
where the plus/minus in $k_1$ and $k_3$ are to be taken the same on each part of
the manifold which we label ${\cal M}_1^{+}$ and ${\cal M}_1^{-}$.
Defining a family of foliations of ${\cal M}_1^+$ and ${\cal M}_1^-$ as in
\eqref{foliations}, we find the 4-web is linearizable only in three cases: (i)
$\alpha = 0$ (for which a parameterization different from
\eqref{consistentManifoldParameterization} must be used), (ii) $\gamma = 0$, and
(iii) $\alpha = \gamma$. We ignore the case $\beta = 0$ since this corresponds
to two uncoupled linear PDEs which are integrable. In any of the above three
cases, the process defining ${\cal M}_1$ is degenerate. The interaction
coefficient for this process is given by
\begin{align*}
 T^{(1)}_{k_1,k_2,k_3,k_4} &= P^{(1)}_{k_1,k_2,k_3,k_4} +
 S^{(1)}_{k_1,k_2,k_3,k_4},\\
 P^{(1)}_{k_1,k_2,k_3,k_4} &= - \frac{\Omega_{k_1-k_3}U_{k_1, k_1-k_3,k_3}
 V_{k_2+k_4,k_2,k_4}}{(\omega_{k_1} -\Omega_{k_1-k_3} -
 \omega_{k_3})(\Omega_{k_2 + k_4} - \Omega_{k_2} - \Omega_{k_4})} -
 \frac{\omega_{k_1-k_2} U_{k_1,k_2,k_1-k_2} U_{k_3+k_4,k_3,k_4}}{(\omega_{k_1} -
 \Omega_{k_2} - \omega_{k_1-k_2})(\omega_{k_3+k_4} - \Omega_{k_4} -
 \omega_{k_3})},\\
 S^{(1)}_{k_1,k_2,k_3,k_4} &= 2 \frac{U_{k_2 + k_3, k_2, k_3}
 U_{k_1,k_4,k_1-k_4}}{\omega_{k_1}-\Omega_{k_4} - \omega_{k_1-k_4}} + 2
 \frac{V_{k_2 + k_4, k_2, k_4} U_{k_1,k_1-k_3,k_3}}{\omega_{k_1} -
 \Omega_{k_1-k_3} - \omega_{k_3}},
\end{align*}
defined on ${\cal M}_1$. We restrict attention to $k_j<0, j=1,2,3,4$, and find
that both $P^{(1)}$ and $S^{(1)}$ are strictly negative for $\mathcal{A}_1 =
\{\alpha<0, \beta>0, \alpha<\gamma<0\}$ and strictly positive for
$\mathcal{A}_2 = \{\alpha>0,\beta<0, \alpha>\gamma>0\}$. It follows that
$T^{(1)}\neq 0$ on both $\mathcal{A}_1$ and $\mathcal{A}_2$. The complement of
$\mathcal{A}_1 \cup \mathcal{A}_2$ gives exactly the three cases mentioned
above: (i) $\alpha = 0$, (ii) $\gamma = 0,$ and (iii) $\alpha = \gamma$, (again
ignoring $\beta = 0$). It follows that fourth-order terms cannot be removed
from the Hamiltonian using a normal form transformation and thus the system
\eqref{eqn:consistent} cannot be integrable except possibly in these three
cases.

Other resonance manifolds appear when attempting to remove fourth-order terms
from $H$. They are defined by
\begin{align}
  {\cal M}_2 &= \{(k_1,k_2,k_3,k_4): k_1 + k_2 = k_3 + k_4 \text{ and } \omega_1 +
 \Omega_2 = \omega_3 + \Omega_4\},\\
 {\cal M}_3 &= \{(k_1,k_2,k_3,k_4): k_1 + k_2 + k_3 = k_4 \text{ and } \omega_1 +
 \Omega_2 + \omega_3 = \Omega_4\},\\
 {\cal M}_4 &= \{(k_1,k_2,k_3,k_4): k_1 + k_2 + k_3 + k_4=0 \text{ and } \omega_1 +
 \Omega_2 + \omega_3 + \Omega_4 = 0\},\\
 {\cal M}_5 &= \{(k_1,k_2,k_3,k_4): k_1 + k_3 = k_2 + k_4 \text{ and } \omega_1 +
 \omega_3 = \Omega_2 + \Omega_4 \}.
\end{align}
The investigation of each manifold results in resonances except in the three
cases (i) $\alpha = 0$, (ii) $\gamma=0$, and (iii) $\alpha = \gamma$. 

The three singled-out systems are:
\begin{align}
  (\text{i})\quad & \alpha = 0: \left\{ \begin{aligned}
  \begin{split} u_t + 2\beta u_x &= -
    2\beta(u v)_x ,\\
    v_t + \beta v_x + \beta v v_x + \gamma v_{xxx} &= -
    \beta(|u|^2)_x .
  \end{split} 
  \end{aligned}\rt.\\
  (\text{ii})\quad & \gamma = 0: \left\{ \begin{aligned}
   \begin{split} u_t + 2\beta u_x + \alpha u_{xxx} &= -
    2\beta(u v)_x ,\\ v_t + \beta v_x + \beta v v_x  &= -
    \beta(|u|^2)_x .
  \end{split} 
  \end{aligned}\rt.\label{newEqns}\\
(\text{iii})\quad& \alpha = \gamma: \left\{ \begin{aligned}
   \begin{split} u_t + 2\beta u_x + \gamma u_{xxx} &= -
    2\beta(u v)_x ,\\
    v_t + \beta v_x + \beta v v_x + \gamma v_{xxx} &= -
    \beta(|u|^2)_x .
  \end{split} 
\end{aligned}\rt.
\end{align}
In order to determine if these systems are integrable, one must look to remove
fourth-order terms from the Hamiltonian. This is not pursued here. The further
investigation of these singled-out systems and their potential physical
relevance is an interesting topic for future study. 

The integrability of coupled KdV equations has been studied extensively (e.g.
\cite{MR2169484, MR1600996, MR1693782, MR1455573}). The systems \eqref{newEqns},
however, have not been shown to be integrable or nonintegrable in any of these
sources. The method employed here is different from the methods in the above
citations, and it may provide additional insight into this well-studied topic.

\section{Conclusion}
Using normal-form theory, we find that the coupled NLS-KdV system
\eqref{eqn:NLS-KdV} is not integrable by either definition in Section 1 since
it has a nondegenerate fourth-order resonance manifold with a nonzero
interaction coefficient function.  We find that the coupled KdV-CKdV system
\eqref{eqn:consistent} is not integrable for the same reasons except
potentially for three choices of parameters: (i) $\alpha = 0,$ (ii) $\alpha =
\gamma$, or (iii) $\gamma = 0$. We cannot verify the integrability of equations
\eqref{newEqns} using the methods described in this paper. In particular, the
tools borrowed from the theory of web geometry cannot be used when looking at
fifth-order resonances and higher since the results on linearizability and rank
are unique to 4-webs. Our methods do however provide a way to isolate
potentially interesting problems and can be used to show the nonintegrability
of other systems of equations, particularly those with complicated four-wave
interactions.

\section*{Acknowledgements}
The authors acknowledge Eugene Benilov, Nghiem Nguyen, and Vladimir Zakharov for
helpful conversations and ideas. The authors thank Takayuki Tsuchida for
bringing to our attention the extensive results on the classification of coupled
KdV equations. This work is generously supported by the National Science
Foundation grant number NSF-DMS-1522677 (JU). Any opinions, findings, and
conclusions or recommendations expressed in this material are those of the
authors and do not necessarily reflect the views of the funding sources.

\begin{appendices}
\section{Web geometry background \label{sec:webGeometry}}
It is sufficient for our purposes to define web geometry \cite{chernWebGeometry}
for 2-dimensional manifolds.
\begin{definition}
    Let $(x,y)$ be local coordinates for a 2D (real) manifold. Then a
    \textit{$d$-web} is the local foliation of the manifold by $d$ curves
    defined by
  \begin{align}
    u_j(x,y) = \text{const}, \quad 1\leq j \leq d,\label{defn:web}
  \end{align}
  where $u_j(x,y)$ are smooth functions.
\end{definition}
We need a definition regarding the geometry of the webs.
\begin{definition}
  A $d$-web is \textit{linearizable} if it is diffeomorphic to a $d$-web formed
  by $d$ one-parameter foliations of straight lines on the plane
  \cite{akivis2005linearizability}. 
\end{definition}

\begin{example}
For a 2D manifold with local coordinates $(x,y)$, the two families of curves
\begin{align}
  x = c_1, \quad y = c_2, \quad \frac xy = c_3, \quad x+y = c_4, 
\end{align}
for $c_1, c_2, c_3, c_4 \in \R$ define a 4-web. The web is linear since the defining
curves are lines. Since the web is linear, it is trivially linearizable.
\end{example}

For our purposes $d$ will always equal 4. An important invariant of a given web
is the \textit{rank} of the web.
\begin{definition}
  The \textit{rank of a $d$-web} is equal to the number of linearly independent
  relations of the form
    \begin{align}
         \sum_{j=1}^d f_j(x,y) = 0.
    \end{align}
\end{definition}
The $4$-webs we work with are always defined on the resonance manifold $M$
\eqref{defn:resonanceManifold} and hence always have rank at least equal to 2
since
\begin{align}
  \sum_{j=1}^4 \sigma_j k_j = 0, \quad \text{ and }\quad \sum_{j=1}^4
  \sigma_j\omega^{(\ell)}(k_j) = 0. \label{defn:manifoldEqns}
\end{align}
The rank of a $4$-web is equal to $2,3$ or $\infty$ \cite{chernWebGeometry}.
Generally infinite-rank webs are disregarded in the theory of web geometry, but
in our application they are possible. The web has infinite rank when the only
solution to \eqref{defn:manifoldEqns} is of the form $k_\alpha = k_\beta$ and
$k_\gamma = k_\delta$ for $(\alpha,\beta,\gamma,\delta) \in \{1,2,3,4\}$.  This
corresponds to so-called billiard scattering \cite{balk}. Therefore as long as
the resonance manifold does not correspond to billiard scattering, the rank of
the web is either 2 or 3. Poincar\'{e}'s Theorem of web geometry states that a
planar 4-web is of rank three if it is linearizable. Since our resonance
manifolds are degenerate if there exists another linearly independent relation
on the manifold, it is sufficient to determine whether or not a 4-web defined on
the manifold is linearizable to determine if it is degenerate. To determine if a
4-web is linearizable, we use the algorithm and Mathematica code developed in
\cite{akivis2005linearizability}.

\end{appendices}

{\footnotesize
\bibliographystyle{siam}
\bibliography{IntegrabilityOfLongAndShortWaves.arxiv}

\begin{thebibliography}{10}

\bibitem{ablowitz_segur-IST}
{\sc M.~J. Ablowitz and H.~Segur}, {\em Solitons and the inverse scattering
  transform}, vol.~4 of SIAM Studies in Applied Mathematics, Society for
  Industrial and Applied Mathematics (SIAM), Philadelphia, Pa., 1981.

\bibitem{akivis2005linearizability}
{\sc M.~A. Akivis, V.~V. Goldberg, and V.~V. Lychagin}, {\em Linearizability of
  d-webs, d{ $\geq$ }4, on two-dimensional manifolds}, Selecta Mathematica, 10
  (2005), pp.~431--451.

\bibitem{MR3129020}
{\sc J.~Albert and S.~Bhattarai}, {\em Existence and stability of a
  two-parameter family of solitary waves for an {NLS}-{K}d{V} system}, Adv.
  Differential Equations, 18 (2013), pp.~1129--1164.

\bibitem{MR2240820}
{\sc J.~Angulo~Pava}, {\em Stability of solitary wave solutions for equations
  of short and long dispersive waves}, Electron. J. Differential Equations,
  (2006), pp.~No. 72, 18.

\bibitem{Fordy_Antonowicz-HamStruct}
{\sc M.~Antonowicz and A.~P. Fordy}, {\em Hamiltonian structure of nonlinear
  evolution equations}, in Soliton theory: a survey of results, Nonlinear Sci.
  Theory Appl., Manchester Univ. Press, Manchester, 1990, pp.~273--312.

\bibitem{balk}
{\sc A.~Balk and E.~Ferapontov}, {\em Invariants of 4-wave interactions},
  Physica D: Nonlinear Phenomena, 65 (1993), pp.~274--288.

\bibitem{benilovPersonal}
{\sc E.~Benilov}.
\newblock personal communication, 2016.

\bibitem{benilov1983integrability}
{\sc E.~Benilov and S.~Burtsev}, {\em To the integrability of the equations
  describing the langmuir-wave-ion-acoustic-wave interaction}, Physics Letters
  A, 98 (1983), pp.~256--258.

\bibitem{MR1681850}
{\sc L.~Chen}, {\em Orbital stability of solitary waves of the nonlinear
  {S}chr\"odinger-{K}d{V} equation}, J. Partial Differential Equations, 12
  (1999), pp.~11--25.

\bibitem{chernWebGeometry}
{\sc S.~S. Chern}, {\em Web geometry}, Bull. Amer. Math. Soc. (N.S.), 6 (1982),
  pp.~1--8.

\bibitem{MR2309177}
{\sc A.~J. Corcho and F.~Linares}, {\em Well-posedness for the
  {S}chr\"odinger-{K}orteweg-de {V}ries system}, Trans. Amer. Math. Soc., 359
  (2007), pp.~4089--4106.

\bibitem{deconinckSegal-LongShort}
{\sc B.~{Deconinck}, N.~V. {Nguyen}, and B.~L. {Segal}}, {\em {The interaction
  of long and short waves in dispersive media}}, Journal of Physics A
  Mathematical General, 49 (2016), p.~415501.

\bibitem{MR2674102}
{\sc J.~Dias, M.~Figueira, and F.~Oliveira}, {\em Well-posedness and existence
  of bound states for a coupled {S}chr\"odinger-g{K}d{V} system}, Nonlinear
  Anal., 73 (2010), pp.~2686--2698.

\bibitem{MR1455573}
{\sc A.~Karasu}, {\em Painlev\'e classification of coupled {K}orteweg-de
  {V}ries systems}, J. Math. Phys., 38 (1997), pp.~3616--3622.

\bibitem{MR1600996}
{\sc I.~V. Kulemin and A.~G. Meshkov}, {\em To the classification of integrable
  systems in {$1+1$} dimensions}, in Symmetry in nonlinear mathematical
  physics, {V}ol. 1, 2 ({K}yiv, 1997), Natl. Acad. Sci. Ukraine, Inst. Math.,
  Kiev, 1997, pp.~115--121.

\bibitem{magri1978Hamiltonian}
{\sc F.~Magri}, {\em A simple model of the integrable hamiltonian equation},
  Journal of Mathematical Physics, 19 (1978), pp.~1156--1162.

\bibitem{MR1693782}
{\sc S.~Y. Sakovich}, {\em Coupled {K}d{V} equations of {H}irota-{S}atsuma
  type}, J. Nonlinear Math. Phys., 6 (1999), pp.~255--262.

\bibitem{zakharov_schulman-degenerativeDispersionLaws}
{\sc E.~I. Schulman and V.~E. Zakharov}, {\em Degenerative dispersion laws,
  motion invariants and kinetic equations}, Phys. D, 1 (1980), pp.~192--202.

\bibitem{MR2169484}
{\sc T.~Tsuchida and T.~Wolf}, {\em Classification of polynomial integrable
  systems of mixed scalar and vector evolution equations. {I}}, J. Phys. A, 38
  (2005), pp.~7691--7733.

\bibitem{Wiggins}
{\sc S.~Wiggins}, {\em Introduction to applied nonlinear dynamical systems and
  chaos}, vol.~2 of Texts in Applied Mathematics, Springer-Verlag, New York,
  1990.

\bibitem{zakharov1974hamiltonianFormalism}
{\sc V.~E. Zakharov}, {\em The hamiltonian formalism for waves in nonlinear
  media having dispersion}, Radiophysics and Quantum Electronics, 17 (1974),
  pp.~326--343.

\bibitem{zakharovSchulmanImportantDevelopmentsInScattering}
{\sc V.~E. Zakharov, A.~Balk, and E.~I. Schulman}, {\em Conservation and
  scattering in nonlinear wave systems}, in Important developments in soliton
  theory, Springer Ser. Nonlinear Dynam., Springer, Berlin, 1993, pp.~375--404.

\bibitem{zakharovSchulmanWhatIsIntegrability}
{\sc V.~E. Zakharov and E.~I. Schulman}, {\em Integrability of nonlinear
  systems and perturbation theory}, in What is integrability?, Springer Ser.
  Nonlinear Dynam., Springer, Berlin, 1991, pp.~185--250.

\end{thebibliography}
}

\end{document}